\documentclass[aps,prl,twocolumn,superscriptaddress,showpacs,floatfix]{revtex4-1}
\usepackage{graphicx,amssymb}
\usepackage[fleqn]{amsmath}
\usepackage{amsfonts}
\usepackage{dcolumn}% Align table columns on decimal point
\usepackage{bm}% bold math
\usepackage{braket}
\usepackage{multirow} 
\usepackage{MnSymbol}

\graphicspath{{images/}}

\renewcommand{\vec}[1]{\mbox{\boldmath $#1$}}

\begin{document}

\date{\today}

\title{Nuclear charge and neutron radii and nuclear matter: trend analysis}

\author{P.-G.~Reinhard}
\affiliation{Institut f\"ur Theoretische Physik, Universit\"at Erlangen, D-91054 Erlangen, Germany}

\author{W. Nazarewicz}
\affiliation{Department of Physics and Astronomy and NSCL/FRIB Laboratory,
Michigan State University, East Lansing, Michigan  48824, USA}
\affiliation{Institute of Theoretical Physics, Faculty of Physics, 
University of Warsaw, Warsaw, Poland}

\begin{abstract}
	\begin{description}
		\item[Background]
			Radii of charge and neutron distributions are fundamental nuclear properties. They depend on both nuclear interaction parameters related to the equation of state of infinite nuclear matter and on quantal shell effects, which are strongly impacted by   the presence of nuclear surface. 
		\item[Purpose]
			In this work, by studying the dependence of charge and neutron radii, and neutron skin, on nuclear matter parameters,  we assess different mechanisms that drive nuclear sizes.
		\item[Method]
			We apply nuclear density functional theory using a family of  Skyrme functionals obtained by means of different optimization protocols targeting specific nuclear properties. By performing the Monte-Carlo sampling of reasonable functionals around the optimal parametrization,  we study correlations between nuclear matter paramaters and observables characterizing charge and neutron distributions of spherical closed-shell nuclei $^{48}$Ca, $^{208}$Pb, and $^{298}$Fl. 
		\item[Results]
			We demonstrate the existence of the strong converse relation  between the nuclear charge radii and the saturation density of symmetric nuclear matter $\rho_0$, and  also between the neutron skins and the slope of the symmetry energy $L$. For functionals optimized to experimental binding energies only, proton and neutron radii are weakly correlated due to canceling trends from different nuclear matter parameters.
		\item[Conclusion]
			We show that by requiring that the nuclear functional reproduces the empirical saturation point of symmetric nuclear matter practically fixes the charge (or proton) radii, and vice versa. This explains the recent results of ab-initio calculations with two-nucleon and three-nucleon forces optimized simultaneously to binding energies and radii of selected nuclei. The neutron skin uncertainty primarily depends on the slope of  the symmetry energy. Consequently, imposing a constraint on both $\rho_0$ and $L$ practically determines the nuclear size, modulo small variations due to shell effects.
	\end{description}
\end{abstract}

\pacs{21.10.Gv, 21.60.Jz, 21.65.Cd, 21.65.Mn}

\maketitle

%
%  INTRODUCTION
%
% General introductionNMP
\textit{Introduction} --  Radii of proton (or charge) and neutron distributions in atomic nuclei are key  observables that can be directly related to fundamental properties of nuclear matter and to the nature of nuclear force (see Ref.~\cite{(Hag16)} and references quoted therein). 
In heavy nuclei, the excess of neutrons gives
rise to a neutron skin, characterized by the neutron distribution
extending beyond the proton distribution. 
The neutron skin  has been found to correlate with
a number of observables in finite nuclei and nuclear and neutron matter \cite{(Ton84),(Rei99),(Bro00),Horowitz_2001,(Hor01a),(Typ01),(Fur02),(Sam09),(Cen09),Naz10a,(Roc11),Piekarewicz_2012,(Agr12),(Lat13),Fatt13,Nazarewicz_2013,(Kor13skins),Roca-Maza_2013,(Meu14),Don15,(Ina15),(Mon15)}; hence, it beautifully links finite nuclei with extended nuclear matter found, e.g.,  in neutron stars.

The goal of this study is to understand the relations between proton and neutron radii, and neutron skins  using nuclear density functional theory (DFT) \cite{(Ben03)}, which is
a tool of choice in microscopic studies of complex
nuclei. In particular, we inspect the relations between nuclear matter parameters characterizing 
effective interactions, here
represented by  Skyrme energy density functionals
(EDFs) adjusted to experimental data using different optimization strategies. 
By means of the statistical
covariance technique, we quantify the intricate relation between the proton and neutron radius, and explain the recent results of a comparative study for $^{48}$Ca ~\cite{(Hag16)}.

%section
\textit{The strategy} -- 
To explore the correlations between neutron radius,
proton radius, and neutron skin, we use
the tools of linear regression based on least-squares ($\chi^2$),
which were adopted recently in the nuclear context in
Refs.~\cite{Naz10a,(Fat11),Reinhard_2013,Reinhard_2013a,Gao13,Dob14a,GorCap14,Erl14b,Kort15,Piek15,RocMaz15,Rei15d}. In particular,  we use here analysis of covariances
(statistical correlations) between observables, error propagation, and
an exploration of $\chi^2$ in the vicinity of the best fit. Our starting point is the  parameterization SV-min \cite{Kluepfel_2009} optimized to the pool of ground-state data. The corresponding  set of fit-observables had been carefully selected to include
only nuclei which have very small correlation corrections
\cite{Kluepfel_2008} and thus can be described reliably within a standard single-reference 
nuclear DFT. Since the  set of fit-observables constraining SV-min contains
also information on radii deduced from the charge form factor data
\cite{Fri82a,Fri86a},  this makes this EDF parametrization  less useful for the present study, whose objective is 
to explore correlations with charge radii. Indeed, one should not trust
correlations for an observable which was included in the fit as the behavior of $\chi^2$ in the direction of this observable is usually very rigid \cite{RocMaz15}. To provide
sufficient leeway to explore radii, the radius information should be excluded from the fit. Thus we consider here the SV-min set of fit observables
excluding the data on radii.
This leaves in the fit pool only energy information, namely  binding energies
of sixty semi-magic nuclei, pairing gaps from odd-even binding energy differences  for long isotopic and isotonic chains, and a few selected spin-orbit
splitting in doubly-magic nuclei. (For details of the fit data, see
Tables III and IV of Ref.~\cite{Kluepfel_2009}.) The EDF optimized to this dataset is referred to as SV-E in the following.
This parametrization and the effect of
omission of radius information had been discussed 
in Refs.~\cite{Erl14b,Rei15d}. Here, we use SV-E merely as a tool to explore radius correlations.

The Skyrme EDF is described by means of  14 parameters. The pairing functional contains
three parameters: proton and neutron pairing strengths and a
parameter defining the density dependence. Two parameters are used for
calibrating the isoscalar and isovector spin-orbit force
\cite{Rei95d}. Two parameters are necessary to tune the surface energy.
Finally, there remain seven parameters characterizing  volume properties.  These
are fully equivalent to key properties of uniform symmetric nuclear matter at
equilibrium, called henceforth Nuclear Matter Parameters (NMP). Those 
are: the saturation density $\rho_0$ and energy-per-nucleon ${E}/{A}$, of symmetric nuclear matter; incompressibility $K$ and effective
mass $m^*/m$ characterizing the isoscalar response; and  symmetry energy $J$,
slope of symmetry energy  $L$, and Thomas-Reiche-Kuhn sum rule enhancement
factor $\kappa_\mathrm{TRK}$ characterizing the isovector response. Those NMP will be used in the following to sort the results and
 establish correlations with radii.

For the following analysis, we employ three strategies.
First, we employ the standard covariance analysis explained, e.g., in
Refs.~\cite{Bra97aB,Dob14a}. Here, we compute the covariance matrix for SV-E  and use it to deduce the covariances (correlations)
between  the observables of interest.  
Second,  we explore explicitly the hyper-surface of
``reasonable parametrizations" in the vicinity of the SV-E parameter set. Recall that, around the minimum of  $\chi^2$  parametrizations are distributed   with probability
$W(\vec{p})\propto\exp{\left[\chi^2(\vec{p})\right]}$ where $\vec{p}$
stands for the (twelve) free parameters of the model and $\vec{p_0}$ is the SV-E parameter set.  We sample this
distribution in a Monte-Carlo fashion by representing it by an
ensemble of 2000 parametrizations. Thereby, we confine the search to the space
of $\vec{p}$ with $W(\vec{p})>1/2$ to avoid excessive amount of
unsuccessful hits in the large  parameter
space. The close vicinity of $\vec{p_0}$ suffices for the present purposes
as it contains all crucial trends and correlations. (For more discussion of such strategy, see Ref.~\cite{GorCap14}.)
%%%%%%%%%%%%%%%%%%%%%%%%%%%
\begin{figure}[htb]	\includegraphics[width=1.0\linewidth]{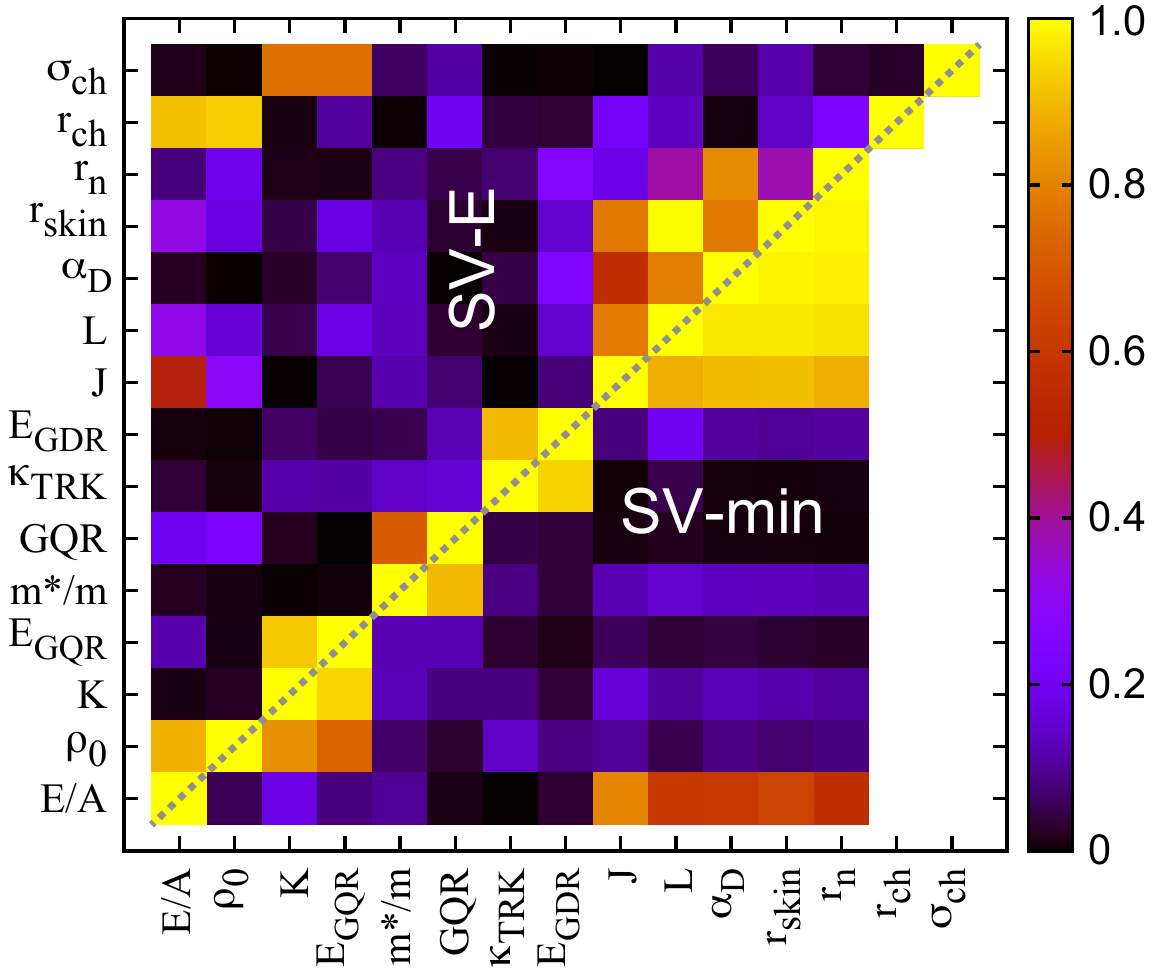}
	\caption{(Color online) \label{fig:alignmatrix-comp} 
Covariance matrices for a selection of observables and NMP computed with SV-min \cite{Kluepfel_2009} (below the diagonal) and 
SV-E \cite{Erl14b,Rei15d} (above the diagonal). Nuclear observables (in $^{208}$Pb) are: charge surface thickness $\sigma_{\rm ch}$; root-mean-square (rms) charge radius $r_{\rm ch}$;
rms neutron radius $r_{\rm n}$; neutron skin $r_{\rm skin}$; electric dipole polarizability $\alpha_{\rm D}$; and giant resonance energies $E_{\rm GDR}$ and $E_{\rm GMR}$. The NMP corresponding to 
the symmetric nuclear matter include: incompressibility
$K$, symmetry energy $J$, symmetry energy slope $L$, isoscalar effective mass $m^*/m$; TRK sum-rule enhancement $\kappa_{\rm TRK}$; and density $\rho_0$ and energy $E/A$ at the saturation point. The correlations with the charge form factor data $\sigma_{\rm ch}$ and $r_{\rm ch}$ are 
not shown for SV-min as these  quantities we included in the correspondins set of fit-observables.
}
\end{figure}
%%%%%%%%%%%%%%%%%%%%%%%%%%
Finally,  we employ the rules of error propagation in the
context of $\chi^2$ fits. We use this to explore the sensitivity of
the radii to NMP by constraining the fit by selected NMP (while using
always exactly the same pool of fit observables) and studying resulting changes in the
uncertainties of the predicted radii. The chosen NMP is always fixed at
the SV-E value. This
means that the optimal parameters $\vec{p_0}$ remain the same. What changes are
allowed variations in $\vec{p}$ which, in turn, impact the
extrapolation uncertainties. We shall see a strong correlation if
one NMP reduces significantly the uncertainty of an observable.

\textit{Results} -- We begin by inspecting in Fig.~\ref{fig:alignmatrix-comp} the covariance matrices of SV-min and SV-E. The general pattern seen in Fig.~\ref{fig:alignmatrix-comp} was discussed in Refs.~\cite{Erl14b,Rei15d}. 
The strong correlations between the isovector indicators 
(symmetry energy $J$, symmetry energy slope $L$, rms neutron radius $r_{\rm n}$, neutron skin $r_{\rm skin}=r_{\rm n}-r_{\rm p}$, and  electric dipole polarizability $\alpha_{\rm D}$) seen in SV-min become significantly degraded in SV-E, with the strongest remaining correlation being that between 
$r_{\rm skin}$ and $L$. Indeed, as concluded in \cite{Erl14b,Rei15d}
$L$ is  the leading bulk parameter for
isovector static response.
The charge radius $r_{\rm ch}$ in SV-E correlates  very well with the saturation point ($\rho_0$ and $E/A$) but rather poorly with other quantities. On the other hand, the neutron radius in SV-E has a reasonable correlation with $\alpha_{\rm D}$ but it is hardly correlated with $r_{\rm ch}$, $\rho_0$, and $E/A$. 

%%%%%%%%%%%%%%%%%%%%%%%%%%
\begin{figure}[htb]	\includegraphics[width=1.0\linewidth]{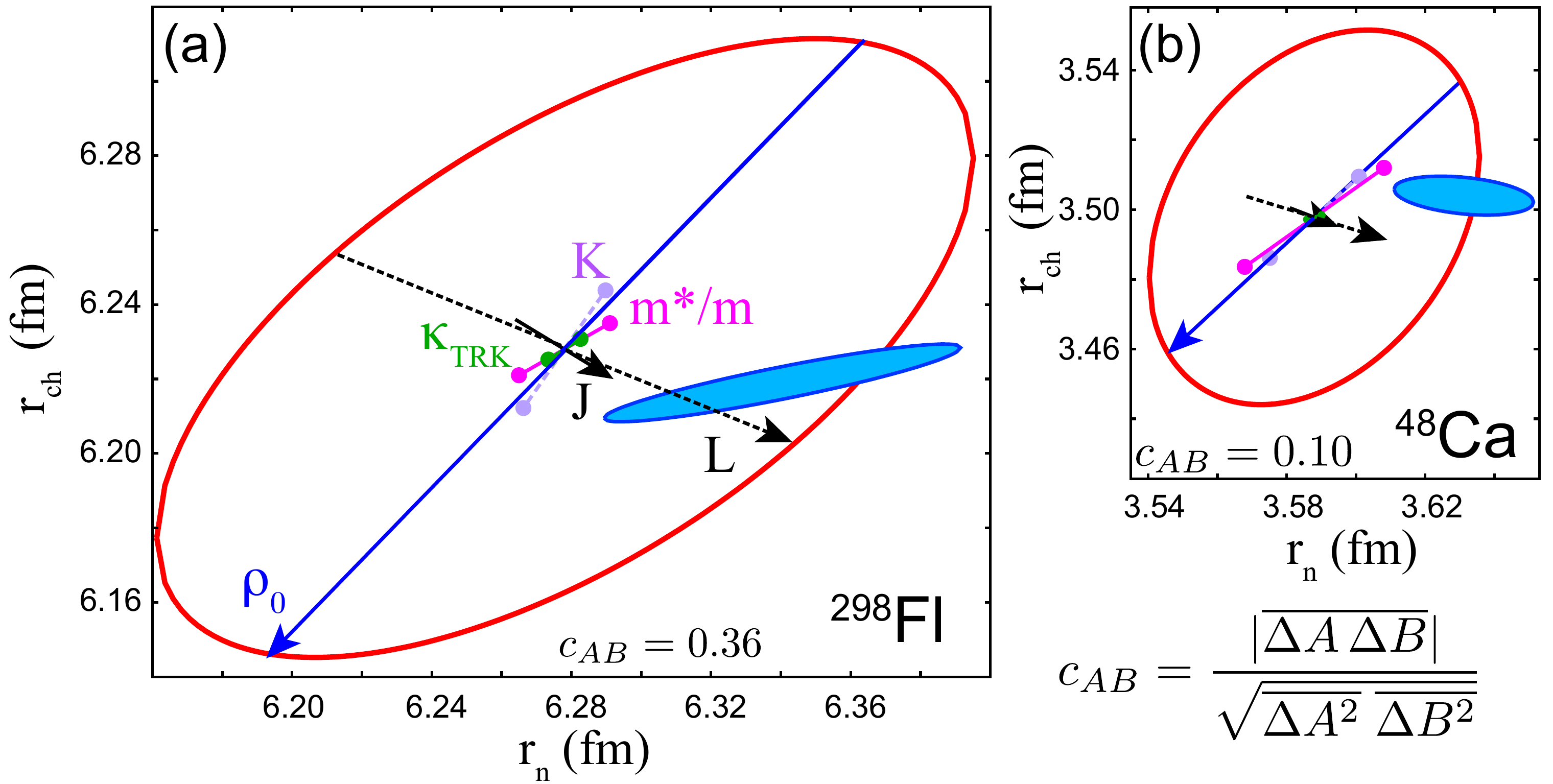}
	\caption{(Color online) \label{fig:ellips} 
Variance ellipsoids with SV-E in the  ($r_{\rm ch}, r_{\rm n}$)-plane  for  $^{298}$Fl (a) and $^{48}$Ca (b).
The arrows/segments indicate the direction of changing radii when varying one NMP as indicated. Their lengths represent the magnitude of corresponding variations. The narrow ellipsoids mark the SV-min results. The correlation coefficient $c_{AB}$ between $r_{\rm ch}$ and $r_{\rm n}$ is indicated in both cases.}
\end{figure}
%%%%%%%%%%%%%%%%%%%%%%%%%%
In the following, we shall test the robustness of the correlations
$r_{\rm ch}\leftrightarrow \rho_0$ and
$r_{\rm skin} \leftrightarrow L$ by inspecting trends in $^{48}$Ca,
$^{208}$Pb, and also in $^{298}$Fl ($Z=114, N=184$) -- a spherical superheavy nucleus, in which the leptodermous expansion is expected to work best \cite{Rei06a}. 
By considering a medium-mass, heavy, and superheavy nucleus, we can assess whether  finite-size (or shell) effects do not cloud our conclusions.
To illustrate the impact of NMP on $r_{\rm ch}$ and $r_{\rm n}$, Fig.~\ref{fig:ellips} shows  the SV-E variance ellipsoids in the  ($r_{\rm ch}, r_{\rm n}$)-plane. Consistent with results displayed in Fig.~\ref{fig:alignmatrix-comp}, the variance ellipsoids are primarily impacted by the variations in the directions of $\rho_0$ and $L$.
The impact of other NMP is much less. Interestingly, the directions of trends due to changes in $\rho_0$ and $L$ (marked by arrows) are fairly different. That is, increasing $\rho_0$ decreases both  $r_{\rm ch}$ and $r_{\rm n}$, as expected from the relation between the density and the Wigner-Seitz radius. On the other hand, increasing $L$ decreases $r_{\rm ch}$ and increases $r_{\rm n}$. The trend due to changes in $J$ generally follows that of $L$, albeit with a much smaller magnitude. Due to the compensating trends, the correlation $r_{\rm ch}\leftrightarrow r_{\rm n}$ is very small; namely, it is $c_{AB}$=0.10 for $^{48}$Ca and it increases to
$c_{AB}$=0.36 for $^{298}$Fl. This illustrates that these two quantities are not strongly coupled by the Skyrme EDF. Figure~\ref{fig:ellips} also shows the corresponding SV-min ellipsoids (narrow, blue). As expected, these are very narrow in the direction of $r_{\rm ch}$, as this quantity has been constrained in the fit of SV-min. On the other hand, the uncertainty in  $r_{\rm n}$ is significant. 

%%%%%%%%%%%%%%%%%%%%%%%%%%
\begin{figure}[htb]	\includegraphics[width=1.0\columnwidth]{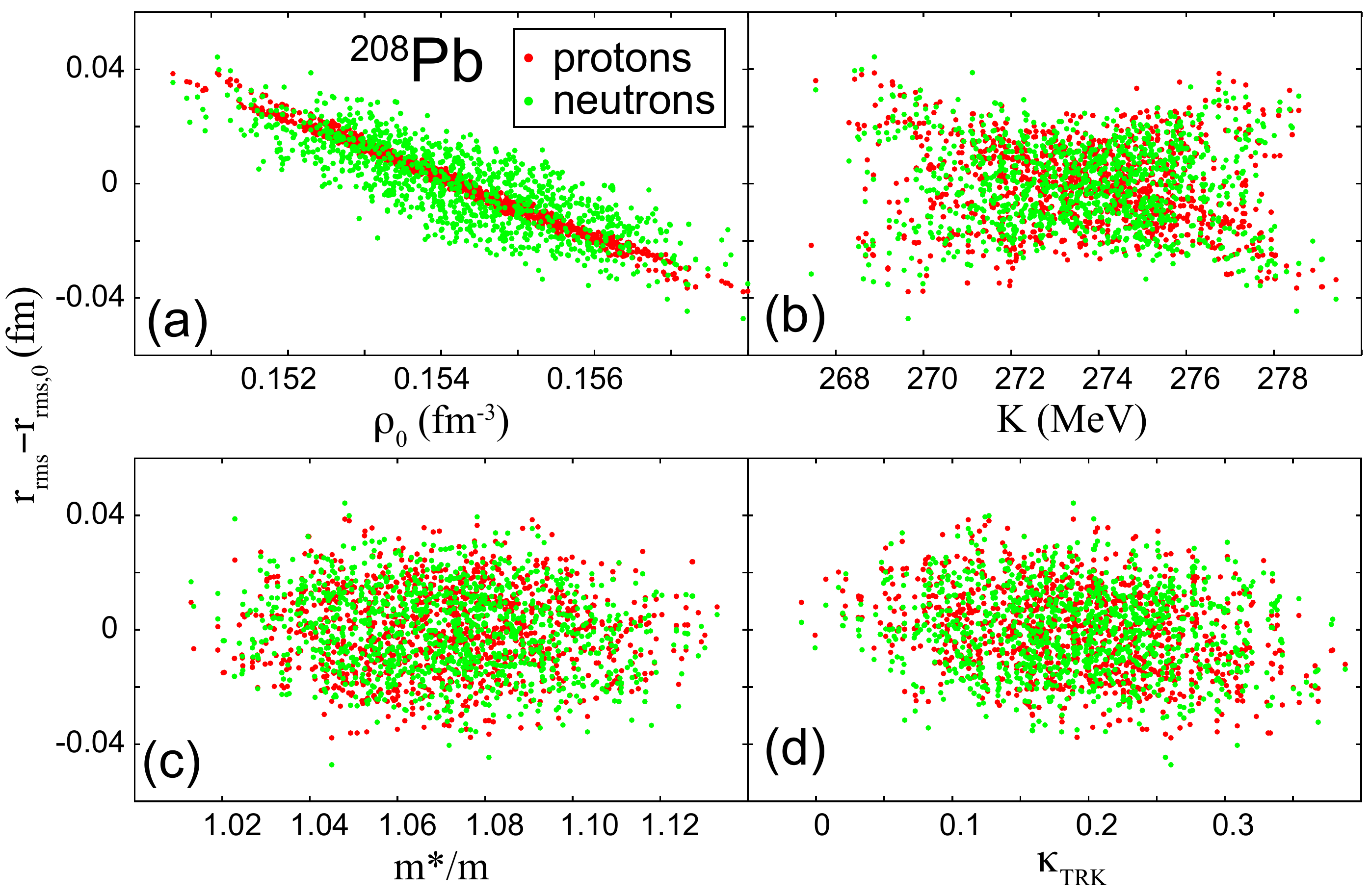}
	\caption{(Color online)  \label{fig:rms-vs-NMP}
	Proton (red), and neutron
  (green) rms radii in $^{208}$Pb with respect to the SV-E values from the ensemble of 2000 parametrizations
in the vicinity of the optimal fit SV-E drawn versus different NMP: $\rho_0$ (a), $K$ (b), $m^*/m$ (c), and $\kappa_{\rm TRK}$ (d).}
\end{figure}
\begin{figure}[htb]	\includegraphics[width=1.0\columnwidth]{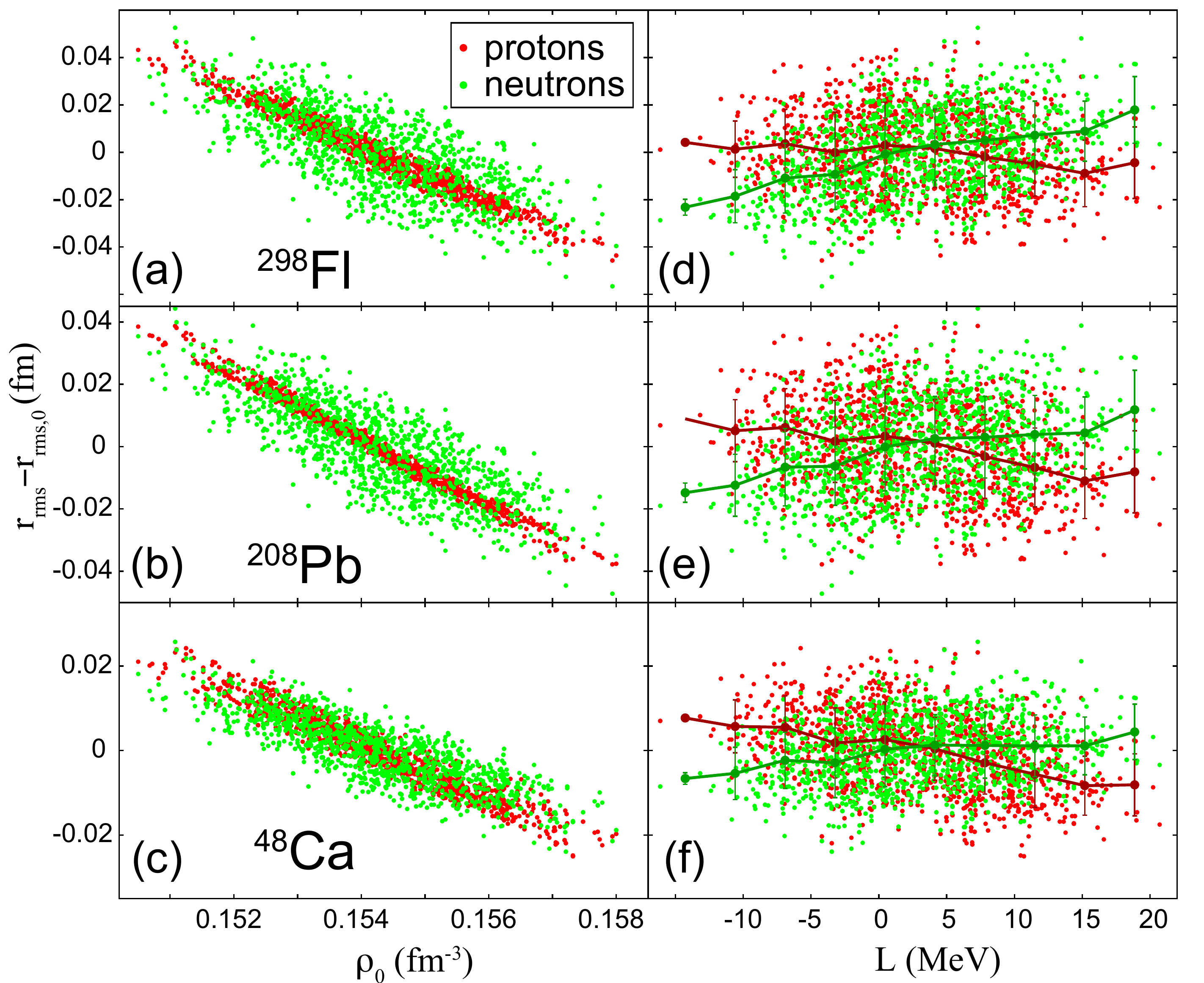}
	\caption{(Color online) \label{fig:rms-vs-L} 
Similar as in Fig.~\ref{fig:rms-vs-NMP} but versus $\rho_0$ (left) and $L$ (right) for $^{298}$Fl (top), $^{208}$Pb (middle), and
$^{48}$Ca (bottom). To illustrate the trends, the right panels show also averages and variances of radii taken
over bins in $L$.}
\end{figure}
\begin{figure}[htb]	\includegraphics[width=1.0\columnwidth]{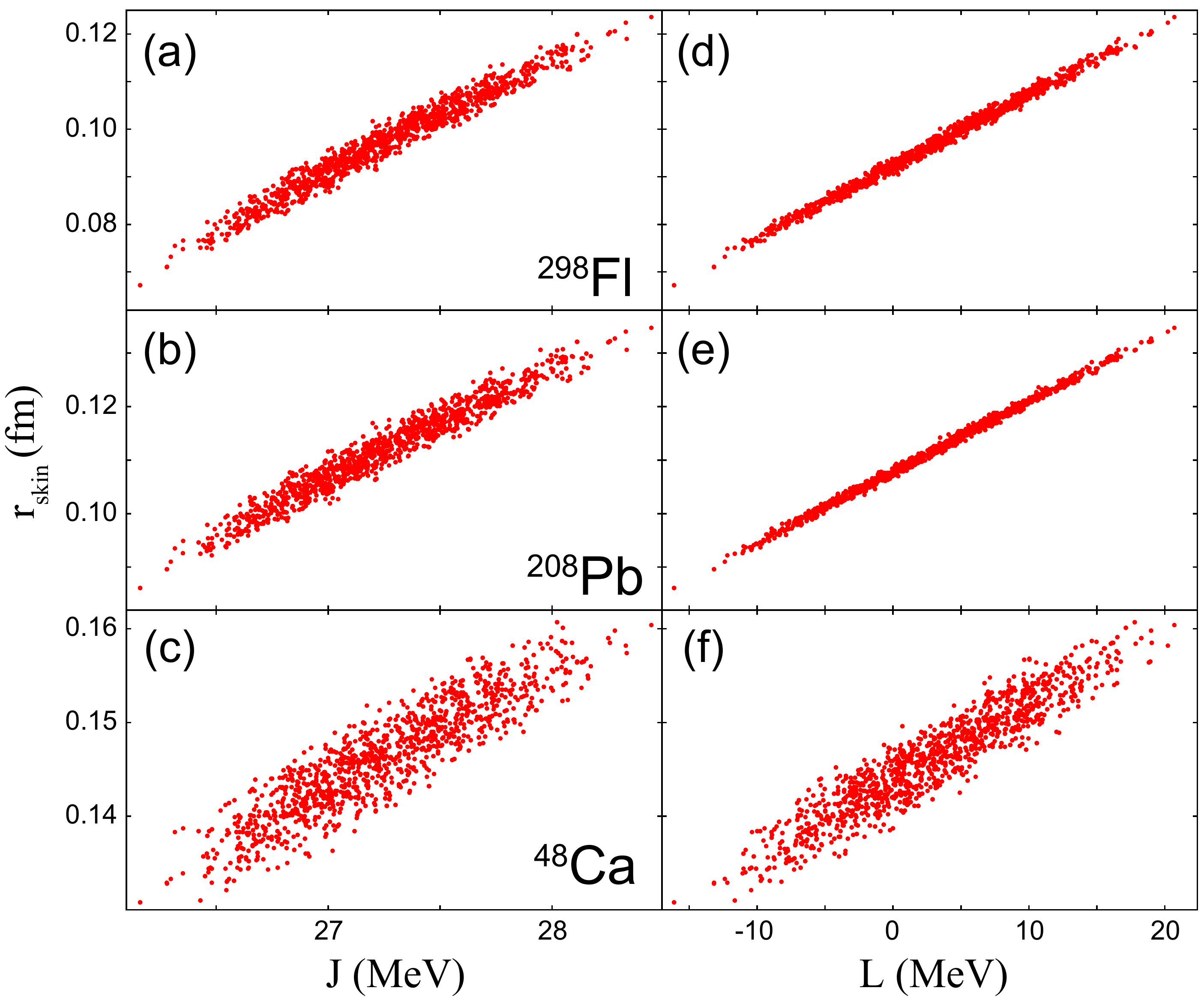}
	\caption{(Color online)  \label{fig:skin-vs-L}
Neutron skins from the ensemble of 2000 parametrizations
in the vicinity of the optimal fit SV-E  versus $J$ (left) and  $L$ (right) for the three nuclei under consideration.
}
\end{figure}

%%%%%%%%%%%%%%%%%%%%%%%%%%
Figures \ref{fig:rms-vs-NMP}-\ref{fig:skin-vs-L} display systematic trends obtained with the
ensemble of 2000 parametrizations around SV-E. These results fully confirm our previous findings. Namely, $\rho_0$, $L$, and $J$ nicely correlate with rms radii and neutron skin while  
$K$, $m^*/m$, and $\kappa_{\rm TRK}$ do not. The behavior of radii in Fig.~\ref{fig:rms-vs-L} is consistent with the trends  in Fig.~\ref{fig:ellips} for the variance ellipsoids. It is interesting to notice that for a fixed value of $\rho_0$, the spread of the proton (or charge) radii is fairly narrow, while it is significantly broader for the neutron radii. Finally, as shown in Fig.~\ref{fig:skin-vs-L},  neutron skins correlate well with $J$ but their correlation with $L$ is superior, especially for heavy nuclei.
%%%%%%%%%%%%%%%%%%%%%%%%%%
%%%%%%%%%%%%%%%%%%%%%%%%%%

%%%%%%%%%%%%%%%%%%%%%%%%%%
%%%%%%%%%%%%%%%%%%%%%%%%%%

\begin{figure}[htb]
\includegraphics[width=1.0\linewidth]{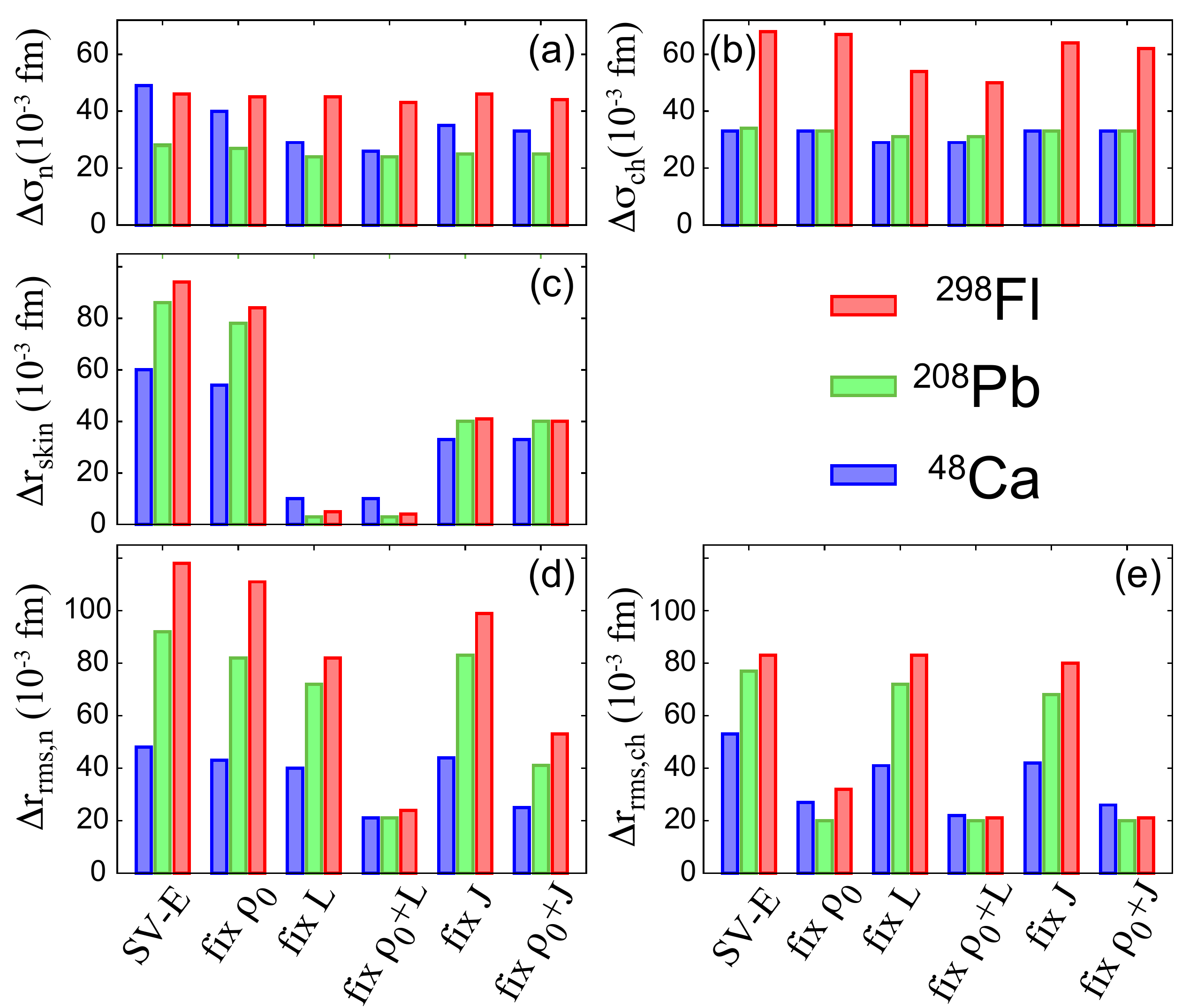}
\caption{\label{fig:variances-fixNMP}
(Color online) Uncertainties in the predictions
  of rms neutron and charge  radii (bottom), neutron skins (middle), and
  surface thicknesses (top) for different EDF
  fits. The reference EDF is SV-E.  Other EDF fits
use the same  pool of fit-observables as SV-E but constrain one
  or two NMP, as indicated, at the SV-E values.
 }
\end{figure}

The strong  $r_{\rm ch}\leftrightarrow \rho_0$ and
$r_{\rm skin} \leftrightarrow L$ relations can be quantified by studying the predicted uncertainties on radii and skins. To this end, we carry out additional EDF optimizations by using the same  pool of fit-observables as SV-E but constraining one or two NMP at the  values given by  SV-E. Figure~\ref{fig:variances-fixNMP}(e)  illustrates the $r_{\rm ch}\leftrightarrow \rho_0$ correspondence: by constraining the saturation density $\rho_0$ the theoretical uncertainty on
$r_{\rm ch}$ is reduced by $\sim$50\%. Even more striking is the result for the neutron skin in Fig.~\ref{fig:variances-fixNMP}(c): constraining $L$ in the EDF optimization practically fixes
$r_{\rm skin}$. The  correlation $r_{\rm skin} \leftrightarrow L$  follows from the leptodermous analysis \cite{(Cen09)}, which shows that $r_{\rm skin} \propto L/J$.

The surface thickness parameters displayed in Figs.~\ref{fig:variances-fixNMP}(a) and (b) are hardly affected by precise knowledge of $\rho_0$, $L$, and $J$. What about the neutron radii? As seen in Fig.~\ref{fig:variances-fixNMP}(d), fixing $\rho_0$ or $L$ helps reducing theoretical uncertainty slightly, but it is simultaneous knowledge of $\rho_0$ and $L$ that helps reducing the error on $r_{\rm n}$. But this can be viewed as  a secondary effect of the $r_{\rm ch}\leftrightarrow \rho_0$ and
$r_{\rm skin} \leftrightarrow L$ relations. Indeed, $r_{\rm n} = r_{\rm n} + r_{\rm skin}$; hence, $\Delta r_{\rm n} = \Delta r_{\rm n} + \Delta r_{\rm skin}$. The uncertainty of the first term is reduced by precise information on $\rho_0$ while the error on the second term is reduced by our knowledge of $L$. 
%%%%%%%%%%%%%%%%%%%%%%%%%%%%%

Figure~\ref{fig:variances-fixNMP} also shows results with EDFc obtained by constraining the symmetry energy $J$. As discussed earlier, the trends due to $J$ follow those triggered by variations in $L$, but they are weaker. This is because our current knowledge of $J$ is much better than that of $L$.
For instance, the values of $J$ are 31$\pm$2\,MeV in SV-min and 27$\pm$2\,MeV in SV-E (a mere 6-7\% error), while the values of $L$ in SV-min and SV-E are 45$\pm$26 MeV and 3$\pm$63\,MeV, respectively (i.e., they are very uncertain).

%%%%%%%%%%%%%%%%%%%%%%%%%%%%%%%%%%%%%%%%%%%%%%%%%%%%%%%%%%%%%%%%%%%%%%%%
%section
\textit{Conclusion} -- By using the statistical tools of linear regression, we studied  radii of neutron and proton distributions  within the Skyrme-DFT framework. The analysis was carried out for the spherical closed-shell nuclei  $^{48}$Ca, $^{208}$Pb, and  $^{298}$Fl, and the results turned out to  weakly depend  on the system considered, i.e., shell effects. The main conclusion of our study is that there exist, at least within the Skyrme-DFT theory, converse relations between radii in finite nuclei and parameters $\rho_0$ and $L$ characterizing the equation of state of uniform nuclear matter: 
 $r_{\rm ch}\leftrightarrow \rho_0$ and
$r_{\rm skin} \leftrightarrow L$. For instance, by including charge radii in a set of fit-observables, as done for the majority of realistic Skyrme EDFs \cite{(Ben03)}, one practically fixes the saturation density $\rho_0$. Indeed, by adding the charge form factor information to the set of fit-observables of SV-E,  one reduces the theoretical uncertainty on $\rho_0$ by a factor of 7  (from $\rho_0=0.1542\pm0.0076$\,fm$^{-3}$ in SV-E to   $\rho_0=0.1611\pm 0.0011$\,fm$^{-3}$ in SV-min).
Recently, a similar conclusion has been reached in ab-initio calculations based on a chiral interaction NNLO$_{\rm sat}$  optimized simultaneously to low-energy nucleon-nucleon scattering data, as well as binding energies and radii of finite nuclei \cite{Eks15a}. Here, the use of data on charge radii was crucial for reproducing  the empirical saturation point of symmetric nuclear matter. 

By inspecting  various, often competing, trends in Fig.~\ref{fig:ellips} one is tempted to conclude that the  $r_{\rm n}\leftrightarrow r_{\rm p}$ relation is fairly complex. Namely, various trends are possible when moving along `a' trajectory in a parameter space $\{\vec{p}\}$. In this respect, we suggest the two directions that are most important are given by the variations in $\rho_0$ and $L$. Our analysis, in particular the results shown in Fig.~\ref{fig:variances-fixNMP} suggest that reducing the uncertainty on $L$ would lead to a dramatic improvement in our knowledge of neutron skins and neutron radii. 
This is consistent with the findings of Ref.~\cite{(Kor13skins)} that
the slope of the symmetry energy $L$ is the single main
contributor to the statistical uncertainty of $r_{\rm skin}$.
Conversely,  by using the precise information on neutron skins (when available) should allow to improve our knowledge of $L$, hence the neutron matter equation of state.

Finally, we conclude that while adding a constraint on experimental charge radii gives rise to EDFs that are of high fidelity with respect to proton radii, the corresponding  model uncertainties on the neutron radii due to our poor knowledge of $L$ are still appreciable~\cite{(Kor13skins)}. This explains the  Skyrme-DFT results in a recent comparative study for $^{48}$Ca \cite{(Hag16)}.

%%%%%%%%%%%%%%%%%%%%%%%%%%%%%%%%%%%%%%%%%%%%%%%%%%%%%%%%%%%%%%%%%%%%%%%%%%%%%%%%%%%%%

\begin{acknowledgments}
This material is
based upon work supported by the U.S.\ Department of Energy, Office of
Science, Office of Nuclear Physics under  award numbers 
DE-SC0013365 (Michigan State University) and   DE-SC0008511 (NUCLEI SciDAC-3 collaboration); by  the German ministry of Science and Technology, grant number 05P12RFFTG;  by the Deutsche Forschungsgemeinschaft, grant number RE 322-14/1; and by Bundesministerium f\"ur Bildung und Forschung (BMBF) under
  contract number 05P09RFFTB.

\end{acknowledgments}

%\bibliographystyle{apsrev4-1}
%\bibliography{rmsradii}
%merlin.mbs apsrev4-1.bst 2010-07-25 4.21a (PWD, AO, DPC) hacked
%Control: key (0)
%Control: author (72) initials jnrlst
%Control: editor formatted (1) identically to author
%Control: production of article title (-1) disabled
%Control: page (0) single
%Control: year (1) truncated
%Control: production of eprint (0) enabled
%

\end{document}